# AN ALTERNATIVE DERIVATION OF EINSTEIN'S DOPPLER SHIFT AND ABERRATION FORMULAE

Jean Reignier [a)]
*Department of Mathematics, Université Libre de Bruxelles, Brussels, Belgium*

I propose an alternative, purely kinematical, derivation of Einstein's Doppler formula. It is valid for periodic signals of any shape that propagate with the velocity of light. The formula is asymptotic in a parameter proportional to the relative variation of the distance source-receiver during one period. As a by-product, I also derive an alternative proof of Einstein's aberration formulae.

## I. INTRODUCTION.

In his first paper on relativity of June 1905, Einstein derives in a most elegant way the relativistic formulae for the relativistic Doppler shift and the aberration (Ref. 1, § 7). He considers a monochromatic source of light located at infinity, at rest in an inertial reference frame K. The light is received by an observer at rest in another inertial frame K' in uniform motion with respect to K. The Lorentz transformation of the electromagnetic fields associated with this plane electromagnetic wave shows that the phase of the wave is a Lorentz invariant. The desired formulae follow straight from this invariance.

In order to fix the notations of the present paper, let me write again these formulae. Let the Lorentz connection between K and K' be given by:

$$
\begin{aligned}
x' &= k (x + v t) & x &= k (x' - v t') \\
y' &= y & y &= y' \\
z' &= z & z &= z' \\
t' &= k (t + v x) & t &= k (t' - v x')
\end{aligned}
\qquad (1)
$$

where $v$ is the relative velocity of the frames in units of the velocity of light ($c = 1$) and $k = (1 - v^2)^{-1/2}$. Let the source be at rest very far away in the $(x, y)$ plane in a direction that makes an angle $\theta$ with the x-axis and let $\nu$ be the frequency of the emitted light. The signal received by an observer located at the origin O of K (or close enough to this origin to detect the source in the same direction $\theta$) is of the form:

$$\Phi = A \sin 2 \pi \nu ( t + x \cos \theta + y \sin \theta + \varphi ) , \qquad (2)$$

with an amplitude A and a constant phase $\varphi$. (Notice that because we use $c = 1$, the numerical values of the frequency and of the wave number are equal and that the wavelength $\lambda$ is numerically equal to $1/\nu$). Just alike, the signal received by an observer located at the origin O' of K' (or close enough to this origin to detect the source in the same direction $\theta'$) is of the form:

$$\Phi' = A' \sin 2 \pi \nu' ( t' + x' \cos \theta' + y' \sin \theta' + \varphi' ) . \qquad (3)$$

As pointed out by Einstein, the Lorentz transformation of the electromagnetic fields implies that the phase of the signal is Lorentz invariant, i.e.:

$$\nu\ (\ t + x \cos \theta + y \sin \theta + \varphi\ ) = \nu'\ (\ t' + x' \cos \theta' + y' \sin \theta' + \varphi'\ )\ , \qquad (4)$$

(modulo $2\pi$) for any t, x, y, and the corresponding t', x', y' defined by (1). This invariance implies $\nu\varphi = \nu'\varphi'$ (modulo $2\pi$) and the Einstein's formulae for the aberration and the Doppler effect:
- aberration:
$$\sin \theta' = k^{-1}\ \sin \theta\ (1 - v \cos \theta)^{-1}\ , \qquad (5)$$
$$\cos \theta' = (\cos \theta - v) / (1 - v \cos \theta)\ , \qquad (6)$$
or vice versa,
$$\sin \theta = k^{-1}\ \sin \theta'\ (1 + v \cos \theta')^{-1}\ , \qquad (5')$$
$$\cos \theta = (\cos \theta' + v) / (1 + v \cos \theta')\ . \qquad (6')$$
- Doppler effect:
$$\nu' = \nu\ k\ (1 - v \cos \theta)\ , \qquad (7)$$
or equivalently,
$$\nu' = \nu\ k^{-1}\ (1 + v \cos \theta')^{-1}\ . \qquad (7')$$

(The equivalence of formulae {(5), (6), (7)} and {(5'), (6'), (7')} is easily checked through simple trigonometric manipulations or even more simply by exchanging the roles of K and K' variables and the sign of the relative velocity, in conformity with the principle of relativity).

This proof is so simple and so elegant that one could wonder why to look for an alternative one. The motivation is that the model doesn't encounter all possible physical situations. For instance, imagine that the source emits a sequence of very short flashes with some time interval T in between, defining a frequency $\nu = 1/T$ (like the signals emitted by a pulsar). This periodic signal (i.e. the succession of flashes) doesn't correspond to a plane wave representation. Does its frequency transform according to the formula given above? Moreover in laboratory experiments aiming to check the Doppler formula, the source is not located at infinity (although at a distance very large compared to the wavelength). Therefore it is at least desirable to get some estimate of the order of magnitude of the correction. One should also remember that the analogous Doppler effect for sound propagation (though different due to the presence of a propagating medium and its non relativistic character) does properly encounter these remarks through a purely kinematical treatment.

The aim of this paper is to answer to these criticisms by considering from a purely kinematical point of view the case of an arbitrary periodic signal emitted by a source of light located at some finite but large distance from the observer, independently of the shape of the signal.

**II. THE DOPPLER PROBLEM.**

Let S be a source of light at rest in an inertial frame K, which emits signals (of any shape) with a periodicity T. A receiver R at rest in K and located at a Cartesian distance L from S receives these signals with the same periodicity T, after a constant time delay equal to L (remember c =1). Let R' be another receiver at rest in an inertial frame K' in relative motion with respect to K. It receives the signals emitted by S after a time delay that depends on its *variable* distance to S. Because of this variation of the distance, the time interval between two successive signals is not strictly constant and one can only define an *approximate* periodicity T'. The Doppler problem is to calculate this approximate period T' as a function of T and of the angle between the direction of the source and the relative motion of the frames *at a degree of approximation* where one neglects the *changes* that occur in distance and angle during one period T.

I use the transformation (1) between the frame K of the source and the frame K' of the receiver. Let ($x_S$, $y_S$, 0) be the fixed Cartesian coordinates of the source in K and let (x'$_S$(t'), y'$_S$(t'), 0) be its time dependant coordinates in K', these two sets being connected by (1). The receiver is bound to the x'-axis, at a point with coordinates (x'$_R$, 0, 0). Let us suppose that the source emits a flash at time $t_e$ . This emission is an 'event' (in the sense of Einstein) with coordinates ($x_e$, $y_e$, $z_e$, $t_e$) = ($x_S$, $y_S$, 0, $t_e$) in K and respectively (x'$_e$, y'$_e$, z'$_e$, t'$_e$) = (x'$_S$(t'$_e$), y'$_S$, 0, t'$_e$) in K'. The reception of the flash by R defines a second 'event' with coordinates (x'$_r$, y'$_r$, z'$_r$, t'$_r$) = (x'$_R$, 0, 0, t'$_r$) in K' (respectively ($x_r$, 0, 0, $t_r$) = ($x_R(t_r)$, 0, 0, $t_r$) in K ). The first step of our calculation is to compute the time interval t'$_r$ – t'$_e$ between these two events (equivalently, to compute the length of the path travelled by the signal when measured in K'). Clearly, the two events are separated by a 'light like' Minkowski's interval. Therefore, in any inertial frame with coordinates system (X, Y, Z, T), one has:

$$(X_r - X_e)^2 + (Y_r - Y_e)^2 + (Z_r - Z_e)^2 - (T_r - T_e)^2 = 0. \qquad (8)$$

In particular, in the rest frame of the receiver, this relation becomes:

$$( x'_R - x'_S (t'_e))^2 + (y'_S)^2 - (t'_r - t'_e)^2 = 0. \qquad (9)$$

Now, because a light signal propagates with velocity c =1 in any inertial frame, (t'$_r$ - t'$_e$) is the distance that the signal has travelled in K' when going from the source S to the receiver R. Therefore, we see that the spatial distance d' travelled by the flash between the source and the receiver, when considered in K', is the Cartesian distance between the points S and R *at the instant of emission of the signal*:

$$d' = [( x'_R - x'_S (t'_e))^2 + (y'_S)^2 ]^{1/2} . \qquad (10)$$

Therefore, the flash reaches the receiver at the instant t'$_r$ (time used in the frame K'):

$$t'_r = t'_e + d' = t'_e + [( x'_R - x'_S (t'_e))^2 + (y'_S)^2 ]^{1/2} . \qquad (11)$$

Let us also note that in the rest frame of the source, relation (8) becomes:

$$(x_R(t_r) - x_S )^2 + (y_S)^2 - (t_r - t_e)^2 = 0 , \qquad (12)$$

so that the distance d travelled by the flash between the source and the receiver, when considered in K , is the Cartesian distance between the points S and R *at the instant of reception of the signal*:

$$d = [( x_R (t_r) - x_S )^2 + (y_S)^2 ]^{1/2} . \qquad (13)$$

When calculating the Doppler effect we only need the distance d', but both distances d and d' will be used later to calculate the aberration.

The scenario of emission is repeated every T second, the periodicity of the source. Let $t_1$ be one of these emission times and let $t_2 = t_1 + T$ be the next one. In the frame K' the corresponding arrival times at the receiver are respectively $t'_{r1} = t'_1 + d'_1$, and $t'_{r2} = t'_2 + d'_2$. The difference between these two instants is the period T' measured by the receiver:

$$T' = t'_{r2} - t'_{r1} = (t'_2 - t'_1) + (d'_2 - d'_1). \qquad (14)$$

The first contribution is given by the Lorentz transformation of the emission times. As it concerns emissions from the same point $x_S$, it simply amounts to a time dilatation effect :

$$(t'_2 - t'_1) = k (t_2 - t_1) = k\, T. \qquad (15)$$

The second contribution corresponds to the change of distances during one period. It must be calculated at the lowest degree of approximation in some parameter ε that characterizes this change. This parameter will appear in a natural way in the calculation.
The relative motion of the source and the receiver gives:

$$x'_S (t'_2) = k(x_S + v\, t_2) = k(x_S + v\, t_1 + vT) = x'_S (t'_1) + kvT. \qquad (16)$$

Substituting in formula (10) for the distance $d'_2$ gives:

$$\begin{aligned} d'_2 &= [(x'_S (t'_1) + kvT - x'_R)^2 + (y'_S)^2]^{1/2} \\ &= [(d'_1)^2 + 2 (x'_S (t'_1) - x'_R)\, kvT + (kvT)^2]^{1/2} \\ &= d'_1 [\, 1 + 2 (x'_S (t'_1) - x'_R)\, kvT / (d'_1)^2 + (kvT)^2 / (d'_1)^2\, ]^{1/2} \qquad (17). \end{aligned}$$

Two parameters appear in a natural way:
i) the angle θ' between the direction of relative motion and the direction of propagation of the flash in the frame K':

$$\cos \theta'_1 = (x'_S (t'_1) - x'_R) / d'_1\,. \qquad (18)$$

ii) a parameter ε that compares the displacement vT of the source during the time interval T to its distance $d'_1$ at the instant of emission:

$$\varepsilon_1 = (kvT) / (d'_1)\,. \qquad (19)$$

With these two parameters, formula (17) becomes:

$$d'_2 = d'_1 [\, 1 + 2\, \varepsilon_1 \cos \theta'_1 + \varepsilon_1^2\, ]^{1/2}\,. \qquad (20)$$

I recall that the Doppler problem amounts to calculate an *approximate* period T', *neglecting the changes* that occur in distance and angle during one period T. It means that we have to expand the answer (20) in powers of the parameter ε, keeping only the terms that will leave an order zero for the answer T' and an order one to indicate the magnitude of the approximation. This is achieved by the limited expansion:

$$d'_2 = d'_1[ 1 + \varepsilon_1 \cos \theta'_1 + O(\varepsilon_1^2)] , \tag{21}$$

which gives for the difference of lengths of the paths of two successive flashes:

$$d'_2 - d'_1 = kvT \cos \theta'_1 + O(\varepsilon_1) . \tag{22}$$

Conversely, a computation of $d'_1$ starting from $d'_2$ will give:

$$d'_2 - d'_1 = kvT \cos \theta'_2 + O(\varepsilon_2) . \tag{22'}$$

Therefore, it is better to drop the indices 1 and 2 in the right hand side, using some intermediary angle $\theta'$ and parameter $\varepsilon$, keeping of course the same degree of approximation (i.e. $\theta' = \theta'_1 + O(\varepsilon_1) = \theta'_2 + O(\varepsilon_2)$, $d' = d'_1 + O(\varepsilon_1) = d'_2 + O(\varepsilon_2)$). Putting the contributions (15) and (22) together, we obtain the Doppler formula expressed with the angle $\theta'$ used in the receiver frame and also a rough estimate of the effect of finite distance between the source and the receiver:

$$T' = kT (1 + v \cos \theta') + O (\varepsilon) . \tag{23}$$

Writing formula (23) in terms of frequencies,

$$\nu' = \nu k^{-1} (1 + v \cos \theta')^{-1} + O (\varepsilon) , \tag{24}$$

we recover the original Einstein's Doppler formula expressed in the frame of the receiver (formula 7'). This formula is now proved to be valid for any type of periodic light signal. The correction for finite distance of the source is estimated to be linear in the ratio $kvT/d'$, i.e. the relative variation of the distance source-receiver during one period.

**THE ABERRATION PROBLEM.**

The trajectory of a light signal is of course a straight line in both coordinates frames K and K'. It makes with the direction of relative motion (i.e. the x-axis) an angle equal to $\theta$ in frame K and to $\theta'$ in frame K'. We have shown that the lengths of the light path in these two frames are given by simple rules (formulas 10 and 13):
- in K', it is the Cartesian distance between the points S and R *at the instant of emission of the signal*:

$$d' = [( x'_R - x'_S (t'_e))^2 + (y'_S)^2 ]^{1/2} = t'_r - t'_e , \tag{10'}$$

- in K, it is the Cartesian distance between the points S and R *at the instant of reception of the signal*:

$$d = [( x_R (t_r) - x_S )^2 + (y_S)^2 ]^{1/2} = t_r - t_e . \tag{13'}$$

The angle $\theta'$ was introduced in formula (18) through the trigonometric relation:

$$\cos \theta' = (x'_S (t'_e) - x'_R ) / d'$$

$$= (x'_S (t'_e) - x'_R ) /( t'_r - t'_e ) . \quad (25)$$

One could of course use alternatively anyone of the equivalent trigonometric relations:

$$\sin \theta' = y'_S / d' = y'_S /( t'_r - t'_e ) , \quad (26)$$
$$\tan \theta' = y'_S / (x'_S (t'_e) - x'_R ). \quad (27)$$

Just in the same way, the corresponding angle θ in the frame K of the source is defined by anyone of the equivalent trigonometric relations:

$$\sin \theta = y_S / d = y_S /( t_r - t_e ) , \quad (28)$$
$$\cos \theta = (x_S - x_R (t_r)) /( t_r - t_e ) , \quad (29)$$
$$\tan \theta = y_S / (x_S - x_R (t_r) ). \quad (30)$$

The coordinates of events in K and K' are connected by the Lorentz transformation (1), so that:

- for the emission: $\quad t_e = k [ t'_e - v x'_S(t'_e) ] \quad (31)$

- for the reception: $\quad t_r = k [ t'_r - v x'_R] \quad (32)$

Therefore:
$$t_r - t_e = k (t'_r - t'_e) + k v [x'_S(t'_e) - x'_R]. \quad (33)$$

Let us introduce this result in $1/\sin \theta$ , as given by formula (28):

$$1/\sin \theta = (t_r - t_e) / y_S = k (t'_r - t'_e) / y_S + k v [x'_S(t'_e) - x'_R]/ y_S ,$$
$$= k / \sin\theta' + k v / \tan \theta' ,$$
$$= k ( 1 + v \cos \theta')/ \sin \theta'. \quad (34)$$

This is exactly Einstein's formula (5') for the aberration. The cos θ formula (6') is derived from (34) through a simple trigonometric manipulation. The inverse formulae (that give θ' in terms of θ) are easily obtained through trigonometric manipulations or by exchanging the roles of the K and K' variables and the sign of the relative velocity, in conformity with the principle of relativity.

---

[a] Electronic mail: jreignie @ ulb.ac.be
[1] A. Einstein, « Zur Elektrodynamik bewegter Körper », Ann. Phys. **17**, 891-921 (1905).